\documentclass{article}       

\usepackage[utf8]{inputenc}
\usepackage{epsfig}
\usepackage{amsmath}
\usepackage{amsfonts}
\usepackage{amssymb}
\usepackage[title]{appendix}%
\usepackage{mathtools}
\usepackage{graphicx}
\usepackage{lipsum}
\usepackage{units}
\usepackage{colordvi}
\usepackage[dvipsnames]{xcolor}
\usepackage{xspace}
\usepackage{acronym}
\usepackage{booktabs}
\usepackage{hyperref}
\hypersetup{
  colorlinks=true,        
  linkcolor=blue,         
  citecolor=cyan,         
}
\usepackage{mathptmx}      
\usepackage{tikz}
\usetikzlibrary{decorations.markings,positioning,decorations.pathreplacing,calligraphy}

\newcommand{\ra}{\rightarrow}
\newcommand{\tn}[1]{\textnormal{#1}}
\newcommand{\dd}{\tn{d}}
\newcommand{\half}{\frac{1}{2}}
\newcommand{\real}{\operatorname{Re}}
\newcommand{\imag}{\operatorname{Im}}

\newcommand{\ipih}{i\pi/h}

\newcommand{\ee}[1]{e^{#1}}

\newcommand{\erf}[0]{\textnormal{erf}}
\newcommand{\erfc}[0]{\textnormal{erfc}}
\newcommand{\erfcx}[0]{\textnormal{erfcx}}
\newcommand{\w}[0]{\textnormal{w}}
\newcommand{\erfi}[0]{\textnormal{erfi}}
\newcommand{\Dawson}[0]{\textnormal{Dawson}}

\newcommand{\deps}[0]{\epsilon_{\tn{DP}}}
\newcommand{\ldeps}[0]{2^{-52}\approx2.220446049250313\cdot10^{-16}}
\newcommand{\sdeps}[0]{2.2\cdot10^{-16}}

\newcommand{\relerr}[0]{\delta_{\tn{rel}}}

\newcommand{\Res}{\mathop{\mathrm{Res}}}

\newcommand{\erflike}[0]{\textsc{erflike}\@\xspace}
\newcommand{\fadpac}[0]{\textsc{Faddeeva package}\@\xspace}
\newcommand{\mpmath}[0]{\texttt{mpmath}\@\xspace}

\newcommand{\secref}[1]{Sec.~\ref{sec:#1}}
\newcommand{\appref}[1]{Appendix~\ref{sec:#1}}
\newcommand{\figref}[1]{Fig.~\ref{fig:#1}}
\renewcommand{\eqref}[1]{Eq.~(\ref{eq:#1})}
\newcommand{\eqsref}[1]{Eqs.~(\ref{eq:#1})}

\newcommand{\etc}[0]{etc.\@\xspace}
\newcommand{\ie}[0]{i.e.\@\xspace}
\newcommand{\eg}[0]{e.g.\@\xspace}

\begin{document}

\title{Evaluation of complex-valued error-like functions by the
  exponentially-convergent trapezoidal rule}

\author{Federico Maria Guercilena}


\date{\today}

\maketitle

\begin{abstract}
    The exponentially convergent trapezoidal rule is applied to a suitable
integral representation of the Faddeeva function to derive a simple formula for
its evaluation. I describe its properties, strategies for maximising its
efficiency, and its coupling with other evaluation methods (asymptotic
expansions and Maclaurin series). From knowledge of the values of the Faddeeva
function, all other complex-valued error-like functions such as $\erf$ and
$\erfc$ can be easily obtained. The resulting algorithm has been implemented in
a publicly-available C/C++ library named \erflike in IEEE double precision
arithmetic, and tested against more widespread evaluation methods based on
Taylor series and continued fractions, as provided by the widely used \fadpac.
It is found that the algorithm presented here and its implementation achieve
better accuracy and a more regular behaviour of the relative error over vast
regions of the complex plane. In terms of speed of evaluation the \erflike
library also outperforms the \fadpac for complex valued arguments, although not for
real-valued ones.
\end{abstract}

\section{Introduction}
\label{sec:introduction}

The error function is ubiquitously found in virtually every scientific field,
with wide-ranging applications in pure mathematics as well as statistics,
applied mathematics, number theory, physics, \etc. It is defined by the integral
\cite[\href{https://dlmf.nist.gov/7.2.E1}{7.2.1}]{DLMF}
\begin{equation}
  \erf(z)=\frac{2}{\sqrt{\pi}}\int_0^z\ee{-t^2}\dd t\,,
\end{equation}
where the argument $z$ is an arbitrary complex number. From this definition
several properties of $\erf$ can be established. In particular, the error
function is entire (\ie holomorphic over the whole complex plane), and it
assumes real values when its argument is real. Being the integral of an
exponential it is manifestly not an elementary function, but a transcendental
one.

There exists a set of functions closely related to $\erf$ which share many of
its properties, in particular each of them is an entire, transcendental
function. These are \cite[\href{https://dlmf.nist.gov/7.2}{7.2}]{DLMF}
\begin{align*}
  \label{eq:erf_like_def}
  &\erfc(z) = 1-\erf(z)&\text{   Complementary error function}\,,\\
  &\erfcx(z) =
    \ee{z^2}\erfc(z)&\text{   Scaled complementary error function}\,,\\
  &\erfi(z) = -i\erf(iz)&\text{   Imaginary error function}\,,\\
  &\Dawson(z) =
    \frac{\sqrt{\pi}}{2}\ee{-z^2}\erfi(z)&\text{   Dawson integral}\,,\\
  &\w(z) = \erfcx(-iz)&\text{ Faddeeva function}\,,
\end{align*}
where despite its name the imaginary error function too assumes real values for
real arguments. In the following, this group of functions is referred to as
\emph{error-like} functions. Other related functions are the Voigt profile and
the Voigt functions \cite[\href{http://dlmf.nist.gov/7.19}{7.19}]{DLMF}.

Error-like functions are non-trivial to evaluate numerically, even though
several alternative representations are known. Being entire functions they equal
their Taylor series at any point in the complex plane, and in particular their
Maclaurin series are well known. Furthermore there exist asymptotic expansions,
representations as continued fractions and several alternative integral
representations as well. Algorithms developed for their evaluation usually rely
on these series expansions, \eg joining a Maclaurin series close to the origin
with a continued fraction far from it. The most well known algorithm for this
purpose seems to be the one by \cite{Gautschi1970}, later improved by
\cite{Poppe1990}. A different algorithm has been proposed in
\cite{Zaghloul2011}, essentially based on writing the Gaussian in the definition
of the error function as a sum of hyperbolic cosines.

These algorithms form the basis for the \fadpac \cite{fadpac}, an open-source
C/C++ code that provides functions to evaluate all error-like functions for
arbitrary complex arguments, with accuracy close to IEEE double-precision
$\deps\approx\sdeps$. Bindings to the \fadpac functions for many popular
programming and scripting languages (in particular R and Matlab/Octave) exists.
Furthermore the Python library scipy and the Julia language rely on the \fadpac
to provide implementations of error-like functions. Therefore, the \fadpac can
be see as a sort of \emph{de facto} standard, and in this work I employ it as
reference point to gauge the performance of the algorithm described here.

This work explores a different approach to the problem of evaluating error-like
functions, based on the application of the composite trapezoidal rule to the
numerical quadrature of an integral representation of the Faddeeva function. The
rationale behind this choice is the observation that the integral in question
satisfies the conditions necessary for the trapezoidal rule to converge
exponentially \cite{Trefethen2014}, yielding a very cheap and yet accurate
evaluation formula. Similar approaches have been considered in various past
works
\cite{Chiarella1968,Matta1971,Hunter1972,Luke1969,Luke1975,vanderLaan1984}, but
none have provided a practical implementation and a complete characterisation of
the speed and accuracy of these schemes.

By applying the technique of \cite{Goodwin1949}, I derive a remarkably simple
formula to evaluate the Faddeeva function. From knowledge of its value, all
other error-like functions can be easily computed. I discuss the practical
considerations involved in employing this evaluation strategy and maximising its
efficiency. The proposed algorithm is then implemented, targeting double
precision accuracy, in an open-source C/C++ library named \erflike. The accuracy
and computational performance of this implementation are then measured and
compared to the \fadpac.

The algorithm proposed here and its implementation efficiently achieve double
precision accuracy, and in vast regions of the complex plane they outperform the
\fadpac in terms of both accuracy (with a much more regular behaviour of the
relative error as the evaluation argument varies) and evaluation speed. The
\fadpac retains an advantage in terms of computational performance in the case
of real arguments. The algorithm based on the trapezoidal rule has however the
theoretical advantage that it does not rely on \emph{ad hoc} fitted
representations as the \fadpac does and can therefore be used also for arbitrary
precision computations. Finally, complementing the trapezoidal-rule algorithm
with asymptotic expansions and Maclaurin series where appropriate (\ie for very
large or very small arguments, respectively) yields further improvements in
efficiency, making the \erflike library a better choice overall than the
\fadpac.

The rest of this paper is organised as follows. \secref{methods} details the
derivation of the trapezoidal-rule based approximation formula for the Faddeeva
function and its implementation; \secref{results} reports the measurement of the
performance and accuracy of the resulting code, comparing with the \fadpac;
\secref{conclusions} is dedicated to discussion and conclusions. Finally, the
theoretical framework behind the exponential convergence of the trapezoidal rule
is summarised in \appref{trapezoidal_formula}.

\section{Applying the trapezoidal rule to the complex-valued Faddeeva function}
\label{sec:methods}

In order to evaluate an error-like function by means of the trapezoidal rule, it
is necessary to work with a suitable integral representation. As it turns out
the Faddeeva function $\w(z)$ admits such a representation (see
\eqref{integral_w} below), so that the trapezoidal rule is immediately
applicable. From the value $\w(z)$ at any point, the values of all other
error-like functions are easily obtained by closed formulas. As such, the rest
of this section is dedicated to developing an evaluation algorithm targeting the
Faddeeva function. Note that this choice simplifies the comparison with the
\fadpac, since it also treats the Faddeeva function as central.

In the following I make extensive use of the following asymptotic expansion of
the Faddeeva function for large arguments
\cite[\href{http://dlmf.nist.gov/7.12.E1}{7.12.1}]{DLMF}:
\begin{equation}
  \label{eq:asymptotic_w}
  \w(z) \sim \frac{i}{\sqrt{\pi}}
  \sum_{n=0}^{+\infty}\frac{\left(\half\right)_n}{z^{2n+1}}\,,
\end{equation}
where $\left(\half\right)_n = \left(\half\right) \left(\half+1\right)
\left(\half+2\right) \dots \left(\half+n-1\right)$ is Pochammer’s symbol (rising
factorial). This expansion is valid in the angular sector
$-\frac{1}{4}\pi<\arg(z)<\frac{5}{4}\pi$. Within this region, the expansion
justifies the simple estimate
\begin{equation}
  \label{eq:est_w}
  |\w(z)| \approx \frac{1}{\sqrt{\pi}|z|}\,,
\end{equation}
accurate for large $|z|$.

\subsection{Conditioning of the Faddeeva function}
\label{sec:conditioning}

Before devising an evaluation algorithm, it is useful to examine the
conditioning of the target function, in order to identify regions that might
pose difficulties. The relative condition number of a function is defined as the
proportionality factor between its relative change and a given relative change
of its argument, and can be estimated by linear approximation. It quantifies the
amplification of errors in the output value due to errors in the input when
evaluating the function in question.

In the case of the complex Faddeeva function, one obtains:
\begin{equation}
  \label{eq:condition}
  \frac{\w(z+\Delta z) - \w(z)}{\w(z)} = C \frac{\Delta z}{z}
  \quad\Rightarrow\quad
  C \approx \frac{z \w'(z)}{\w(z)} = \frac{2iz}{\sqrt{\pi}\w(z)}-2z^2\,.
\end{equation}
where $C$ denotes the condition number, $\Delta z$ is an arbitrary displacement
and $\w'(z)$ indicates the derivative of $\w(z)$. The last equality results from
the identity $\w'(z)=2i/\sqrt{\pi}-2z\w(z)$
\cite[\href{http://dlmf.nist.gov/7.10.E2}{7.10.2}]{DLMF}. Note that the Faddeeva
function is entire, so this estimate is valid on the whole complex plane.

In the angular sector $-\frac{1}{4}\pi<\arg(z)<\frac{5}{4}\pi$, $|\w(z)|$
attains a maximum of $1$ at $z=0$, and otherwise decreases asymptotically as
$\sim 1/\sqrt{\pi}|z|$ as noted above. As such $|C|$ is never larger than
$\sim1.44$ and tends towards $1$ for large $|z|$. The exceptions to this
behaviour occur at the zeros of $\w(z)$, where $C$ diverges\footnote{The zeros
of $\w(z)$ asymptotically approach the lines $\arg{z}=-\frac{1}{4}\pi$ and
$\arg{z}=\frac{5}{4}\pi$, \ie they are just inside the region of validity of the
asymptotic expansion \eqref{asymptotic_w} \cite{Fettis1973}.}. Outside this
angular sector we have $\w(z)\sim 2\ee{-z^2}$, resulting in $|C|\eqsim 2|z|^2$.

Summarising, we can expect to be able to accurately evaluate $\w(z)$ without
particular difficulties if $-\frac{1}{4}\pi<\arg(z)<\frac{5}{4}\pi$, except in
the vicinity of a zero; while for
$-\frac{3}{4}\pi\leq\arg(z)\leq-\frac{1}{4}\pi$ achieving the desired accuracy
will become increasingly difficult as $|z|$ grows.

\subsection{Trapezoidal rule formula}
\label{sec:w_formula}

As mentioned, the Faddeeva function admits the following integral representation
\cite[\href{http://dlmf.nist.gov/7.7.E2}{7.7.2}]{DLMF}:
\begin{equation}
  \label{eq:integral_w}
  \w(z) = \frac{iz}{\pi}\int_{-\infty}^{+\infty}
  \frac{\ee{-t^2}}{z^2 - t^2} \dd t\,.
\end{equation}
This expression is valid for $\imag(z) > 0$, but also for real $z$ by
interpreting the integral as a Cauchy principal part. This representation has
the same form as \eqref{f_integral} with $K(t;z)=iz/[\pi(z^2 - t^2)]$, so that
formula \eqref{trapezoidal} can be applied to estimate the integral via the
composite trapezoidal rule (see \secref{trapezoidal_formula}). This requires
knowledge of the poles of $K$: in this case two simple poles located at $t=\pm
z$. The resulting formula is singular for particular values of $z$, and to
handle this complication it is useful to consider its staggered version too,
resulting from \eqref{trapezoidals}. The residues at the poles of $K$ necessary
to apply the formulas are easily evaluated as:
\begin{align*}
  \Res\left[\frac{iz\ee{-t^2}} {\pi(z^2 - t^2) (1-\ee{-2\pi i
  t/h})}\right]_{t=\pm z} &= \pm\frac{\ee{-z^2}}{2\pi i(1-\ee{\mp 2\pi i
                            z/h})}\,,\\
  \Res\left[\frac{iz\ee{-t^2}} {\pi(z^2 - t^2) (1+\ee{-2\pi i
  t/h})}\right]_{t=\pm z} &= \pm\frac{\ee{-z^2}}{2\pi i(1+\ee{\mp 2\pi i
                            z/h})}\,,\\
  \Res\left[\frac{iz\ee{-t^2}} {\pi (z^2 - t^2)}\right]_{t=\pm z} &=
                              \mp\frac{\ee{-z^2}}{2\pi i}\,.
\end{align*}

Putting all together, the result is the following formula:
\begin{equation}
  \label{eq:trapezoidal_w}
  \w(z) =
  \begin{dcases}
    \frac{ih}{\pi z} + \frac{2ihz}{\pi}\sum_{n=1}^{+\infty}
    \frac{\ee{-n^2h^2}}{z^2-n^2h^2} + \frac{P\ee{-z^2}}{1-\ee{-2\pi i z/h}}\\
    \frac{2ihz}{\pi}\sum_{n=1}^{+\infty}
    \frac{\ee{-(n-\half)^2h^2}}{z^2-(n-\half)^2h^2} + \frac{P\ee{-z^2}}
    {1+\ee{-2\pi iz/h}}
  \end{dcases}
  - E(h;z)\,,
\end{equation}
where $h$ is the distance between the quadrature nodes and the constant $P$
assumes the values:
\begin{equation}
  \label{eq:poles_P}
  P = \begin{dcases}
    2 & \text{if }\imag(z)<\frac{\pi}{h}\,,\\
    1 & \text{if }\imag(z)=\frac{\pi}{h}\,,\\
    0 & \text{otherwise}\,.
  \end{dcases}
\end{equation}

Both cases in \eqref{trapezoidal_w} are singular for particular real values of
$z$, different between the two. The formula on the second line is the staggered
version of the first (see \appref{trapezoidal_formula}). Following
\cite{Hunter1972}, the first line is employed if $0.25 \leq
\tn{frac}[|\real(z)|/h]\leq 0.75$ and the second otherwise, where
$\tn{frac}(x)=x-\lfloor x \rfloor$ represents the fractional part. This ensures
that the denominator of the sums in \eqref{trapezoidal_w} never vanishes, while
the absolute value of the denominator of the poles contribution, $1\pm\ee{-2\pi
i z/h}$, never attains values lower than $\sqrt{2}$.

The error term $E(h;z)$ in \eqref{trapezoidal_w} reads (see \eqref{Ehz})
\begin{align}
  \label{eq:Ehz_w}
  E(h;z) = 2\ee{-\pi^2/h^2}\frac{iz}{\pi}\int_{-\infty}^{+\infty}
  \frac{\ee{-y^2}}{z^2-(y-i\frac{\pi}{h})^2}\dd y\,,
\end{align}
and clearly vanishes exponentially as $h$ decreases.

As stated at the beginning of the section, this formulas applies in the upper
part of the complex plane. To obtain $\w(z)$ when $\imag(z)<0$, one leverages
the following identity \cite[\href{http://dlmf.nist.gov/7.4.E3}{7.4.3}]{DLMF}:
\begin{equation}
  \label{eq:negative_i}
  \w(-z) = 2\ee{-z^2} - \w(z)\,.
\end{equation}

\subsection{Double precision implementation}
\label{sec:implementation}

Once a target accuracy $\epsilon$ is chosen, the value of $h$ that allows to
evaluate \eqref{trapezoidal_w} to this accuracy with minimal computational
effort can be computed, along with the number of terms after which the infinite
series can be truncated. In the following I discuss these choices (and other
considerations important to achieve good performance) targeting IEEE double
precision, \ie so that the relative error satisfies:
\begin{equation}
  \label{eq:relerr}
  \relerr =
  \left|\frac{\w_{\tn{exact}}-\w_{\tn{trapezoidal}}}{\w_{\tn{exact}}}\right| \leq
  \deps = \ldeps\,.
\end{equation}

The algorithm described in the following for the evaluation of the error-like
functions has been implemented in a software library named \erflike, which is
publicly available at
\cite[\href{https://doi.org/10.5281/zenodo.11261631}{DOI:10.5281/zenodo.11261631}]{erflikezenodo}.
The main component of the library is written in C/C++ in a single-file,
header-only fashion. It also provides bindings for the Python language written
with the help of Cython \cite{Behnel2010,cython}. Finally, an unoptimised but
fully functional arbitrary precision implementation, written with the help of
the \mpmath library \cite{mpmath} is included too.

\subsubsection{Choice of $h$ and truncation of infinite sums}
\label{sec:h_choice}

Since by definition $|E(h;z)|$ is the absolute error of \eqref{trapezoidal_w},
in order to determine a suitable value for $h$ it is necessary to require
$\relerr=|E(h;z)|/|w(z)|\leq \deps$. This inequality must hold over the upper
complex plane, where \eqref{trapezoidal_w} is to be applied. Clearly $h$ is most
strongly constrained when $|\w(z)|$ is small, \ie for large $|z|$. Therefore one
can employ \eqref{est_w} to estimate the magnitude of $\w(z)$, while the
expression for $E(h;z)$ can be simplified according to \eqref{Ehz_approx} (note
that $K(t;z)$ monotonically decreases for large $t$, so this approximation is
very close to the correct value). One arrives at the condition
\begin{equation}
  \label{eq:h_estimate}
  \left|\frac{E(h;z)}{\w(z)}\right|\approx2\ee{-\pi^2/h^2}
  \frac{|z|^2}{|z^2+\pi^2/h^2|}\approx 2\ee{-\pi^2/h^2}\leq\deps
  \quad\Rightarrow\quad
  h\leq\frac{\pi}{\sqrt{\log\left(\frac{2}{\deps}\right)}}\,,
\end{equation}
where the second approximate equality results from the assumption of large
$|z|$. The approximations introduced to find this formula tend to underestimate
the correct value of $h$, especially for a low target accuracy. An empirically
determined correction, replacing the value of $h$ obtained from
\eqref{h_estimate} by $h\ra h(1-0.06h)$, solves this issue. Applying the
correction, the value of $h$ necessary to target double precision and adopted in
the following is $h\approx0.5022$.

Having determined $h$, the infinite sums in \eqref{trapezoidal_w} need to be
truncated at some finite number $N$ of terms. When considering the first line of
\eqref{trapezoidal_w}, the requirement $\relerr\leq\deps$ translates to
\begin{equation}
  \frac{1}{|\w(z)|}\left|\frac{2izh}{\pi}\frac{\ee{-N^2h^2}}{z^2-N^2h^2}\right|
  \approx
  \frac{2h\ee{-N^2h^2}}{\sqrt{\pi}}\frac{|z|^2}{|z^2-N^2h^2|}
  \approx
  \frac{2h\ee{-N^2h^2}}{\sqrt{\pi}}
  \leq\deps\,,
\end{equation}
where again the approximate equalities stem from using \eqref{est_w} for large
$|z|$, since the small $|z|$ region constrains $N$ less severely. This results
in
\begin{equation}
  \label{eq:N_estimate}
  N=
  \left\lceil
    \left[\frac{1}{h^2}\log\left(\frac{2h}{\sqrt{\pi}\deps}\right)\right]^\half
  \right\rceil=12\,,
\end{equation}

Applying this line of reasoning to the staggered version of
\eqref{trapezoidal_w} one finds that $N$ or $N+1$ terms should be retained,
depending on the value of $h$ and the target precision. However in all tests
conducted, using the value of $N$ found for the unstaggered case in the
staggered case as well allowed recovering the desired double precision accuracy.
Accordingly, in the following the value of $N$ equals 12 in both cases.

\subsubsection{Partition of the complex plane}
\label{sec:regions}

\begin{figure}[h]
  \begin{center}
    \includegraphics[width=\textwidth]{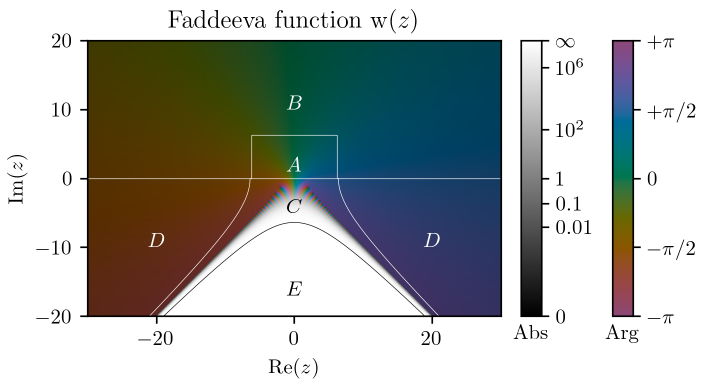}
  \end{center}
  \caption{Domain colouring of the Faddeeva function on the complex plane.
    Superimposed in white and black are the boundaries of the regions of
    \secref{implementation}, themselves indicated by capital letters.}
  \label{fig:regions}
\end{figure}

While formulas \eqref{trapezoidal_w} and \eqref{negative_i} are applicable for
any argument $z$ in the complex plane, depending on its value not all terms in
those equations are necessary to reach the desired accuracy. To achieve maximum
efficiency is therefore beneficial to partition the complex plane in
non-overlapping regions, see \figref{regions}, and apply different strategies of
evaluation in each.

In the upper part of the complex plane, two regions emerge naturally depending
on the possibility to neglect the poles contribution in \eqref{trapezoidal_w}.
Region $A$ is bounded by $0\leq\imag(z)\leq\pi/h$ and $|\real(z)|\leq 6.17$. In
this region no term in \eqref{trapezoidal_w} can be neglected without
sacrificing accuracy. Region $B$ is defined by $\imag(z)\geq 0$, excluding the
points contained in region $A$. Here the poles contribution in
\eqref{trapezoidal_w} either vanishes (see \eqref{poles_P}) or the real part of
$z$ is sufficiently large to neglect it without losing accuracy. It is
sufficient to check the real part of $z$ because $\ee{-z^2}/(1\pm\ee{-2\pi i
z/h})$ decreases rapidly as $\imag(z)$ increases from $0$ towards $\pi/h$. The
condition for neglecting this term is
\begin{equation}
  \frac{2\ee{-\real(z)^2}}{\sqrt{2}\,|\w[\real(z)]|}\leq\deps
  \quad\Rightarrow\quad
  \real(z)^2\geq
  \log\left(\frac{\sqrt{\pi}\,|\real(z)|}{\sqrt{2}\deps}\right)\,,
\end{equation}
  where again \eqref{est_w} has been employed, and the factor of $\sqrt{2}$ in
the denominator is the minimum value attained by $|1\pm\ee{-2\pi i z/h}|$, see
\secref{w_formula}. In the case of double precision, this is equivalent to
$|\real(z)|>6.17$, which justifies the definition of region $A$ above.

If instead $\imag(z)$ is negative, \eqref{negative_i} is used to evaluate
$\w(z)$ from the value of $\w(-z)$. The Gaussian term in \eqref{negative_i}
allows to divide this portion of the complex plane in three regions, in which
the Gaussian is dominant, negligible or neither. These regions are bounded by
the level curves of the complex Gaussian function
$\real(-z^2)=\imag(z)^2-\real(z)^2=\pm g$, corresponding to
$|\ee{-z^2}|=g^{\pm1}$ for some $g>0$, and are labeled in \figref{regions} as
Region $D$ ($\real(-z^2)<g$), Region $E$ ($\real(-z^2)>g$) and Region $C$ (in
between the previous two). The value of the constant $g$ can be determined \eg
by requiring the relative error in neglecting the Gaussian term in
\eqref{negative_i} to be smaller than $\deps$, and specialising this condition
to $\imag(z)=0$. The condition can then be written as
\begin{equation}
  2\sqrt{\pi}\ee{-\real(z)^2}|z|<\deps
  \quad\Rightarrow\quad
  \real(z)^2\geq
  \log\left(\frac{2\sqrt{\pi}\,|\real(z)|}{\deps}\right)\,,
\end{equation}
from which follows that when targeting double precision the appropriate value
for $g$ is $\approx41.024025$.

Finally, note that for purely imaginary arguments $z=ix$ with $x$ real, the
Faddeeva function equals the real-valued scaled complementary error function
$\erfcx(x)$. For purely real arguments instead, $\w(x)$ is in general complex,
but its real part is simply $\ee{-x^2}$. It is therefore beneficial to
specialise the implementation on the axes of the complex plane to two
real-valued functions of real argument, namely $\erfcx(x)$ and $\imag(\w)(x)$.
This allows to employ faster real floating point operations, with computational
advantages that propagate to the computation of other error-like functions as
described in the next section.

\subsection{$\erf$ and other error like-functions}
\label{sec:other_functions}

Once the value of the Faddeeva function is known for some argument $z$, the
value of the error function itself or any of the error-like functions is
computed from the following closed relations:
\begin{align}
  \label{eq:erf_like_comp}
  \begin{split}
  \erfcx(z) &= \w(iz)\\
  \erfc(z) &= \begin{dcases}
    \ee{-z^2}\w(iz)&\text{ if }\real(z)\geq0\\
    2-\ee{-z^2}\w(-iz)&\text{ if }\real(z)<0
  \end{dcases}\\
  \erf(z) &= \begin{dcases}
    1-\ee{-z^2}\w(iz)&\text{ if }\real(z)\geq0\\
    \ee{-z^2}\w(-iz)-1&\text{ if }\real(z)<0
  \end{dcases}\\
  \erfi(z) &= \begin{dcases}
    -i+i\ee{z^2}\w(-z)&\text{ if }\imag(z)\leq0\\
    -i\ee{z^2}\w(z)+i&\text{ if }\imag(z)>0
  \end{dcases}\\
  \Dawson(z) &= \begin{dcases}
    i\frac{\sqrt{\pi}}{2}\left[\ee{-z^2}-\w(z)\right]&\text{ if }\real(z)\geq0\\
    i\frac{\sqrt{\pi}}{2}\left[\w(-z)-\ee{-z^2}\right]&\text{ if }\real(z)<0
  \end{dcases}
  \end{split}
\end{align}
Employing different formulas depending on the sign of the real part of $z$ (or
the imaginary part in the case of \erfi) allows to avoid loss of accuracy due to
multiplying or adding very large and small numbers. Note that the \fadpac
employs the same technique.

\section{Accuracy and performance results}
\label{sec:results}

This section collects and presents the measurements of the accuracy and speed of
the algorithms presented above and implemented in the \erflike library. As basis
of comparison, the corresponding measurements relative to the \fadpac
are employed.

\subsection{Accuracy}
\label{sec:accuracy}

\begin{figure}[h]
  \begin{center}
    \includegraphics[width=\textwidth]{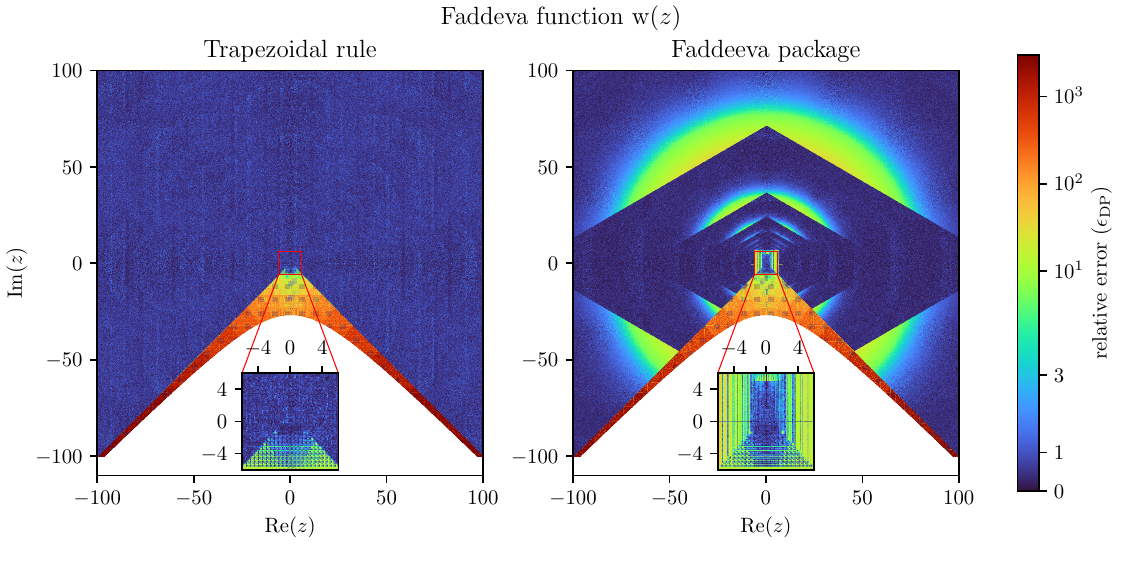}
  \end{center}
  \caption{Relative error in the computation of the Faddeeva function $\w(z)$
    over the complex plane in units of the double precision epsilon $\deps$. The
    colour scale is linear up to $5\deps$ and logarithmic beyond.}
  \label{fig:accuracy_w}
\end{figure}

\figref{accuracy_w} compares the accuracy of the implementation of the algorithm
presented here against the ones implemented in the \fadpac over a wide region of
the complex plane centered around the origin. The insets in both panels magnify
the square region $(-6,6)\times(-6,6)i$. The plotted quantity is the relative
error $\relerr$ (\eqref{relerr}), expressed in units of the IEEE double
precision epsilon, $\deps=\ldeps$. Here and in the following, the reference
values of $\w(z)$ have been obtained by evaluating the expression
$\w(z)=\ee{-z^2}\erfc(-iz)$ in 236-bit arithmetic ($\approx70$ decimal digits)
using the arbitrary-precision Python \mpmath library \cite{mpmath}.

Firstly one can notice that the results summarized in \figref{accuracy_w} match
the behaviour expected by examining the condition number of the Faddeeva
function (see \secref{conditioning}): namely, double precision accuracy is
attained in the angular sector $-\frac{1}{4}\pi<\arg(z)<\frac{5}{4}\pi$, while
outside of it the relative error tends to be higher and grows rather rapidly as
$|z|$ increases. This behaviour can be observed for both the \erflike library
and the \fadpac. Note that in the region
$\imag(z)<0\cap|\imag(z)-\real(z)|\gtrapprox26$ the absolute value of the
Faaddeva function overflows and the result cannot be represented in IEEE double
precision, so the relative error is undefined.

Besides these general considerations, it is apparent that where the Faddeeva
function is well-conditioned the relative error of the trapezoidal rule is
consistently closer to $\deps$ than the \fadpac. The right panel of
\figref{accuracy_w} shows overlapping circle- and rhombus-shaped regions of
growing size, where the accuracy of the \fadpac oscillates between $\unit[\sim
1]{\deps}$ and $\unit[\sim 10^2]{\deps}$. This behaviour seems to be a
manifestation of the variable number of terms that is retained in the continued
fraction representation of $w(z)$ for large $|z|$ \cite{Poppe1990}, so that when
the algorithm switches to a lower-order representation as $|z|$ attains larger
values, significant loss of accuracy occurs. The \erflike library instead
essentially employs a single representation throughout the domain, and is
therefore completely free from these artefacts.

Rather surprisingly, the accuracy of the \fadpac degrades significantly also in
the square-shaped region centred on the origin and shown in the insets of
\figref{accuracy_w}. Here the \fadpac relative error is small very close to the
origin, but increases to $\unit[\sim 10]{\deps}$ when $|\real(z)|\gtrapprox 2$
or $\imag(z)\gtrapprox 5$. Since in many applications the region where
error-like functions are evaluated most often is around the origin, relying on
the \fadpac might result in a noticeable loss of accuracy. The reason for this
behaviour is unclear, but the formulas employed by the \fadpac in this region
appear to be more complex and have more terms \cite{Zaghloul2011} than the ones
resulting from applying the trapezoidal rule. This might increase the occurrence
of cancellation error. On the other hand the potential for cancellation error to
occur seems to be much lower in the case of the \erflike library, which shows a
much more regular and satisfactory behaviour. Its relative error does not differ
significantly from $1\deps$ in this region too, save where the Faddeeva function
itself is ill-conditioned. These observation are confirmed at a more
quantitative level by comparing the average relative error (computed via a
simple arithmetic mean) of the two implementations over the region shown in the
insets: for the \erflike library
$\langle\relerr\rangle\approx\unit[1.84]{\deps}$, while for the \fadpac
$\langle\relerr\rangle\approx\unit[8.91]{\deps}$.

While the analysis above is focused on the Faddeeva function, the same
qualitative behaviour has been observed for each of the error-like functions.
This is to be expected, since both the \erflike implementation and the \fadpac
compute them by starting from the corresponding value of the Faddeeva function
and then applying the same identities (\eqsref{erf_like_comp}).

\begin{figure}[h]
  \begin{center}
    \includegraphics[width=\textwidth]{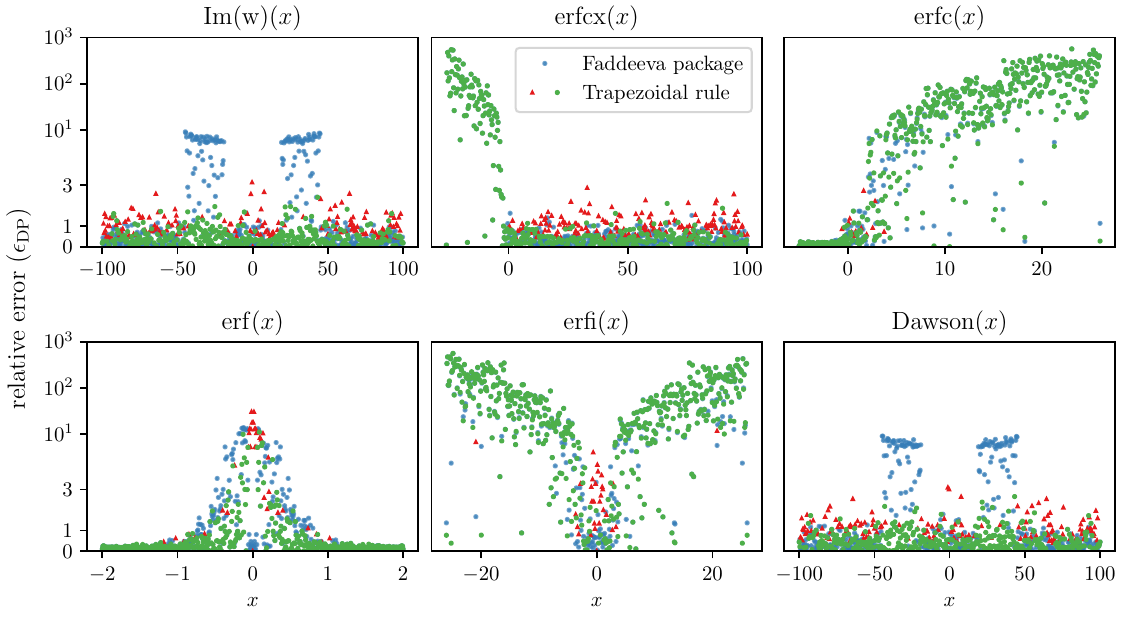}
  \end{center}
  \caption{Relative error in the computation of the error-like functions over
    the real axis in units of the double precision epsilon $\deps$. The ordinate
    scale is linear up to $5\epsilon$ and logarithmic beyond. Values relative to the
    \fadpac are marked by blue dots. Values relative to the \erflike library are
    denoted by green dots if their accuracy is no more than twice worse than that of
    the \fadpac, by red upwards-pointing triangles otherwise.}
  \label{fig:accuracy_re}
\end{figure}

Since error-like functions are often evaluated along the real line, it is
important to investigate the accuracy of the present implementation in this
particular case too. This is shown in \figref{accuracy_re}, in which all
error-like functions are considered. Note that $\imag(\w)$ and Dawson integral
are proportional to each other and as such their plots are extremely similar.
The latter is included only for completeness. It can be seen that in most cases
the \erflike library and the \fadpac behave very similarly, and although at any
given point and for any given function any of the two could be more accurate
than the other, there is no clear overall trend.

Two exceptions to this general observation are visible in \figref{accuracy_re}.
First of all, around $|x|\sim30$ for $\Dawson$ and $\imag(\w)$, the relative
error of the \fadpac grows noticeably, reaching $\unit[\sim10]{\deps}$. This
might be an effect related to switching from a continued fraction representation
for large $|x|$ and a Chebyshev interpolation closer to the origin.
Interestingly, this does not happen in the case of $\erfcx$ or $\erfc$, which
seems to suggest this is a simple oversight in the implementation rather than an
intrinsic property of the algorithm. The relative error of the \erflike library
implementation instead remains consistently below $\unit[\sim 3]{\deps}$.

Secondly, in the case of the odd functions $\imag(\w)$, $\erf$, $\erfi$ and
$\Dawson$, the trapezoidal rule implementation suffers from a slight loss of
accuracy close to the origin. This is not surprising since the origin is a zero
of these functions, which makes them ill-conditioned in its vicinity. An
effective way to mitigate accuracy loss in this cases is simply to evaluate the
relevant Maclaurin series. This is indeed the algorithm implemented in the
\fadpac, resulting in a better accuracy as seen in \figref{accuracy_re}. While
the results shown here focus on characterising the behaviour of the trapezoidal
rule algorithm, the \erflike library offers to the user the possibility of
switching from evaluating \eqref{trapezoidal_w} to evaluating the Maclaurin
series for all real-valued error-like functions at compile time. With this
precaution the relative error drops to the same values observed for the \fadpac.

\subsection{Speed of evaluation}
\label{sec:speed}

The evaluation speed of the \erflike library and the \fadpac has been measured
with the help of the \texttt{nanobench} library \cite{nanobench}. The code has
been compiled using the \texttt{gcc} compiler \cite{gcc}, version 14.1.1, with
flags \texttt{-Ofast -march native}. All tests have been run on an Intel Core
i7-7700HQ CPU running at a frequency of 2.80 GHz.

\begin{figure}[h]
  \begin{center}
    \includegraphics[width=\textwidth]{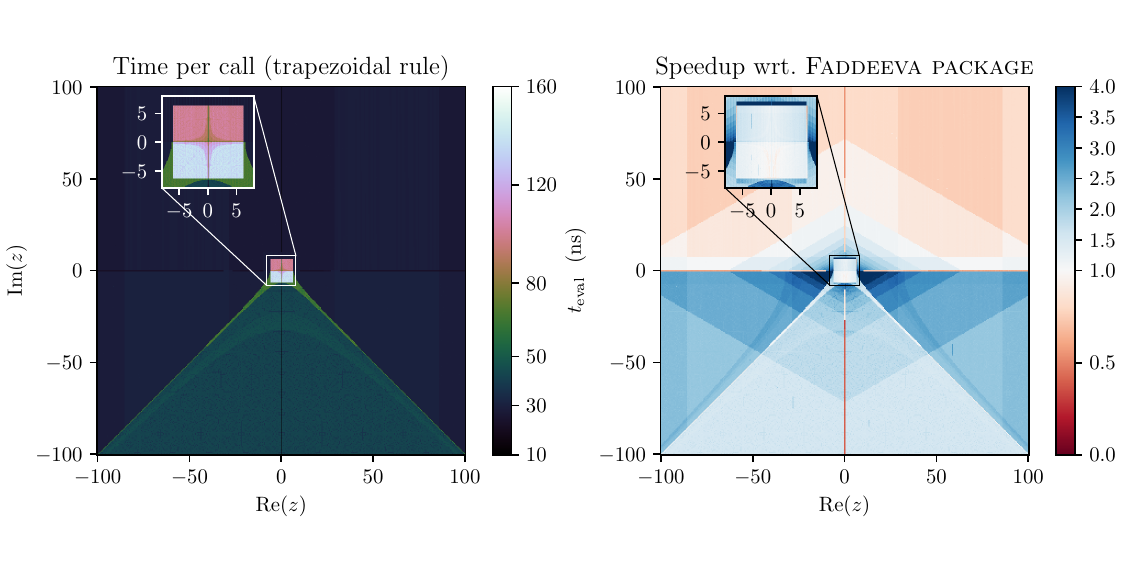}
    \caption{Left panel: time of evaluation of the Faddeeva function $\w(z)$
      over the complex plane for the \erflike implementation, in nanoseconds. Right
      panel: ratio of evaluation speed of the \erflike library with respect to the
      \fadpac.}
    \label{fig:speed_w}
  \end{center}
\end{figure}

The performance in terms of speed of evaluation of the Faddeeva function for the
\erflike library over the complex plane is summarised in \figref{speed_w}. Its
left panel shows the evaluation time in nanoseconds while the right panel
displays the relative speedup (or slowdown) with respect to the \fadpac. The
left panel in particular shows that the behaviour of the evaluation speed of the
\erflike library divides the complex plane in five regions, essentially
characterised by how many exponentials need to be computed for each of them. In
the angular sector $-\frac{1}{4}\pi<\arg(z)<\frac{5}{4}\pi$ and far from the
origin no exponential term is needed and the time of evaluation is around
$\unit[\sim30]{ns}$. In regions $E$ and $C$ defined in \secref{regions} one
exponential term appears, raising evaluation times to $\unit[\sim50]{ns}$ and
$\unit[\sim70]{ns}$ respectively (recall however that in most of region $E$
$\w(z)$ overflows, so high evaluation times would not be particularly
concerning). In region $A$ the presence of the pole contribution in
\eqref{trapezoidal_w}, needing the evaluation of two complex exponentials,
pushes evaluation times up to about $\unit[\sim100]{ns}$. Finally, the region
symmetric to region $A$ across the real axis requires one further exponential
evaluation, reaching the highest evaluation times of about $\unit[\sim150]{ns}$.
Note that the selection of the staggered or unstaggered version of formula
\eqref{trapezoidal_w} depending on the value of the real part of $z$
influences the evaluation time only to a negligible level.

The right panel of \figref{speed_w} reveals that over large portions of the
complex plane the \erflike implementation offers clear advantages over the
\fadpac. Firstly a significant speedup is observed for $\imag(z)<0$, where the
speedup can be up to $\sim4$ times close to the origin. The \fadpac is clearly
penalised by its reliance on a continued fraction representation with a variable
number of terms, resulting in rhombus-like structures as in \figref{accuracy_w}.
The \erflike library instead computes the value $\w(-z)$ and reflects it to the
lower part of the complex plane by means of \eqref{negative_i}, with a clear
improvement in performance. This advantage in performance is present (although
not as striking) in the upper part of the complex plane too, at least for
$|z|\lessapprox25$.

Closer to the origin, in the region $(6.17,-6.17)\times(\pi/h,-\pi/h)i$, the
speed of the \erflike library is marginally higher than that of the \fadpac, as
clearly seen in the insets of \figref{speed_w}). In this region both algorithms
require the evaluation of complex exponentials, resulting in comparable
evaluation times. Finally, in the upper part of the complex plane, the \fadpac
continued fraction representation edges over the \erflike library in terms of
speed, although generally by less than a factor of 2.

\begin{figure}[h]
  \begin{center}
    \includegraphics[width=\textwidth]{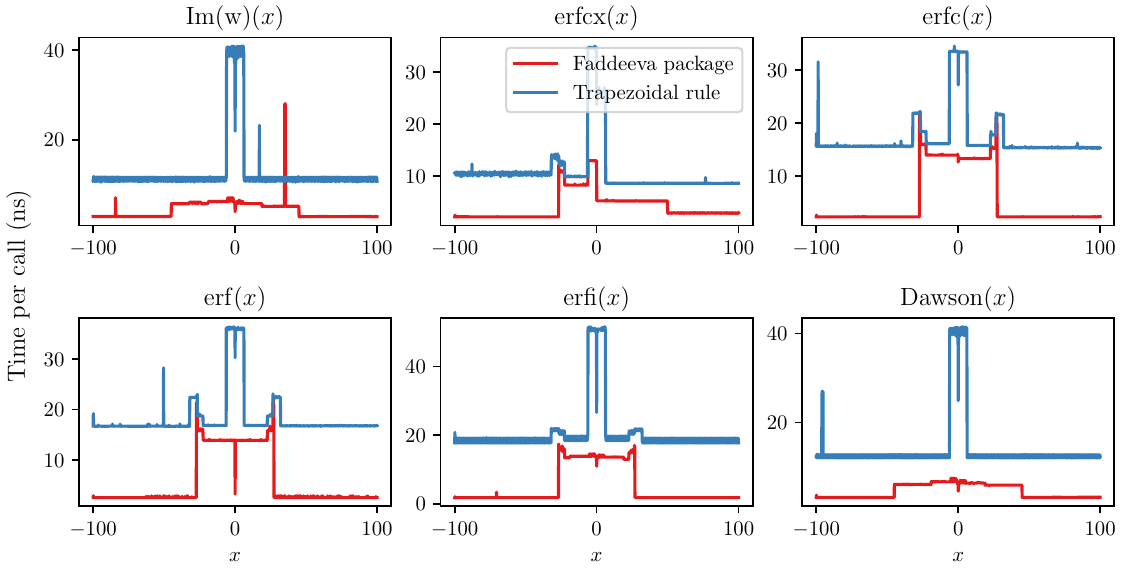}
  \end{center}
  \caption{Evaluation time of the real-valued error-like functions, expressed
    in nanoseconds, for both the \fadpac and \erflike library.}
  \label{fig:speed_re}
\end{figure}

\figref{speed_re} focuses on the evaluation speed in the case of real arguments,
where both the \fadpac and the trapezoidal rule implementations switch from real
to complex arithmetic. As can be seen the \fadpac clearly performs better, being
often more than twice as fast. This is particularly evident around the origin
for any of the error-like functions, where the poles contribution to the
trapezoidal rule require the evaluation of exponentials which severely slows
down the code, up to $\unit[\sim50]{ns}$ per call in the case of $\erfi$.

The \fadpac instead achieves evaluation times smaller than $\unit[10]{ns}$ per
call for every function and most values of the argument, rising slightly close
to the origin. It achieves this performance by employing only a few terms of the
continued fraction representation of these functions far from the origin. Closer
to the origin instead the continued fraction converges very slowly, and the
\fadpac switches to a piecewise fit in terms of Chebyshev polynomials, thus
reducing the operation count and sidestepping the evaluation of quotients and
exponentials, which instead appear prominently in \eqref{trapezoidal_w}. This
explains the difference in performance.

Note however that the Chebyshev representation employed by the \fadpac has been
computed \emph{ad hoc} to maximise speed in the case of double precision, and
its coefficients cannot be determined from closed formulas in more general
cases. \eqref{trapezoidal_w} on which the \erflike library is based is instead
applicable to any required level of accuracy, as detailed in
\secref{implementation}.

Finally, the same compile-time flag mentioned at the end of \secref{accuracy}
allows the user of the \erflike library to switch to the asymptotic expansion
\eqref{asymptotic_w}, truncated to six terms, for both complex- and real-valued
functions when the absolute value of the argument is larger than 30. This
choices preserves the accuracy summarised in \figref{accuracy_w} and
\figref{accuracy_re}, but greatly improves evaluation times far from the origin
(especially for the complex-valued functions), surpassing the performance of the
\fadpac.

\section{Conclusions}
\label{sec:conclusions}

This work is concerned with the accurate and efficient evaluation of the
complex-valued error function and related functions such as $\erfc$, $\erfi$ and
the Faddeeva function. The evaluation strategy focuses on the application of the
trapezoidal rule, including pole contributions, to a suitable integral
representation of the Faddeeva function. Exploiting the exponential convergence
of the trapezoidal rule to the integral in question, I have derived a remarkably
compact evaluation formula.

Next, considerations regarding the efficient use and implementation of this
formula have been discussed, namely the optimal number of terms of the
formula to retain and the possibility to neglect some parts of it in specific
regions of the complex plane without compromising accuracy. Moreover, the
application of the formula to the evaluation to all error-like functions besides
the Faddeeva function has been described.

I have implemented this algorithm in a C/C++ library named \erflike targeting an
accuracy of evaluation equal to IEEE double precision, and measured both its
relative error against reference values of the error-like functions and its
evaluation speed. As a basis of comparison to characterise these results and
place them in the existing literature, the implementation provided by the
\fadpac, which can be seen as a \emph{de facto} standard for the evaluation of
complex error functions, has been employed.

The comparison has revealed that the trapezoidal-rule based algorithm and its
implementation is accurate, displaying a relative error very close to IEEE
double precision level over all regions of the complex plane where the target
functions are well conditioned. This behaviour is much more regular than the one
of the \fadpac, which suffers from significant loss of accuracy in particular
regions. In terms of speed of evaluation too the comparison is positive towards
the trapezoidal-rule algorithm, which is faster than the \fadpac over most of
the complex plane, although slower for real-valued arguments. Coupling this
algorithm with an asymptotic expansion and Maclaurin series where appropriate
further enhances its performance, making it generally the preferable choice.

In conclusion, the application of the exponentially convergent trapezoidal rule
yields an evaluation algorithm for the error-like function which is superior to
the methods widely employed. The \erflike library which implements this
algorithm is publicly available at
\cite[\href{https://doi.org/10.5281/zenodo.11261631}{DOI:10.5281/zenodo.11261631}]{erflikezenodo},
making it possible to reproduce the results presented here. Finally, being
readily available, self-contained and similar in design to the \fadpac, it may
also be widely adopted.


\section*{Conflict of interest}
The author declares that he has no conflict of interest.

\appendix

\section{Composite trapezoidal rule formula}
\label{sec:trapezoidal_formula}

The composite trapezoidal rule formula for the evaluation of definite integrals
on the real line can be written as
\begin{equation}
  \int_{-\infty}^{+\infty}f(x)\dd x = h\sum_{n=-\infty}^{+\infty}f(nh) + E(h;z)\,,
\end{equation}
where $E(h;z)$ is an error term. For generic functions the formula converges to
the value of the integral to second order in $h$, \ie $|E(h;z)|=O(h^2)$. It is
well known however that in particular cases the error term convergences
exponentially. For integrals on the real line, the conditions for exponential
convergence can be stated in terms of fast decay of the integrand as $|x|\ra
\infty$ and analyticity in a strip of the complex plane surrounding the real
line \cite{Trefethen2014}. Even if $f(x)$ is not analytic but meromorphic in
this strip, exponential convergence can still be recovered by taking into
account the poles' contribution. \cite{Goodwin1949} has provided a clear and
succinct way of deriving trapezoidal-rule formulas for a particular class of
integrals and estimate the magnitude of $E(h;z)$. I reproduce his treatment here
in the interest of a self-contained discussion, including the contribution from
the possible poles of the integrand.

Consider a function (in general a complex-valued function of complex variable)
denoted by $f(z)$, which can be represented by an integral of the form
\begin{equation}
  \label{eq:f_integral}
  f(z) = \int_{-\infty}^{+\infty} K(t;z)\,\ee{-t^2}\,\dd t\,,
\end{equation}
where the integration is over the real axis and the function $K$ is even with
respect to $t$ (a non-zero odd part of $K$ would not contribute to the
integral). Let $h$ be a positive real number and assume that $K$ is meromorphic
over the strip $|\imag(t)|<\pi/h$.

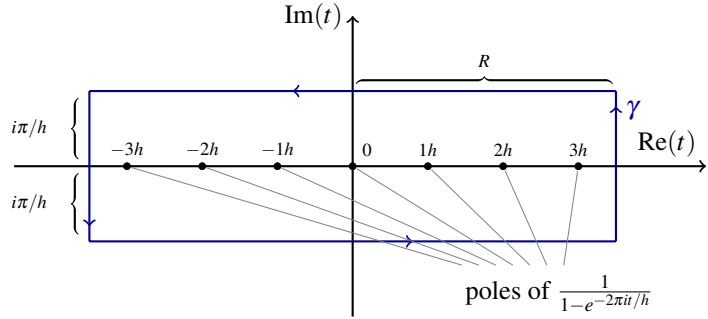
\begin{figure}[h]
  \begin{center}
    \begin{tikzpicture}[thick]
      \draw [->] (-4.5,0) -- (4.7,0) node [above left] {$\real(t)$};
      \draw [->] (0,-2) -- (0,2) node [left] {$\imag(t)$};

      \draw[blue!60!black,decoration={markings,mark=between positions 0.1 and 1
step 0.25 with \arrow{>}},postaction={decorate}] (3.5,-1) -- (3.5,1) node [below
right] {$\gamma$} (3.5,1) -- (-3.5,1) (-3.5,1) -- (-3.5,-1) (-3.5,-1) --
(3.5,-1);

      \node (pih) at (-4.3, +0.5) [font=\scriptsize] {$\ipih$};
      \node (mih) at (-4.3, -0.5) [font=\scriptsize] {$\ipih$};
      \node (R) at (1.75, +1.4) [font=\scriptsize] {$R$};
      
      \draw [decorate, decoration={calligraphic brace}] (-3.65,+0.1) -- (-3.65,+0.9);
      \draw [decorate, decoration={calligraphic brace}] (-3.65,-0.9) -- (-3.65,-0.1);
      \draw [decorate, decoration={calligraphic brace}] (0.05,+1.1) -- (3.45,+1.1);

      \foreach \n in {-3,...,-1,1,2,...,3}{%
        \draw[fill] (\n, 0) circle (1pt) node [above,font=\scriptsize] {$\n h$};}
      \draw[fill] (0, 0) circle (1pt) node [above right,font=\scriptsize] {$0$};

      \node (hpoles) at (2.75, -1.7) {poles of $\frac{1}{1-\ee{-2\pi i t/h}}$};
      \draw[fill]
      (0, 0) coordinate [circle,fill,inner sep=1pt] (p0)
      (1, 0) coordinate [circle,fill,inner sep=1pt] (pp1)
      (-1, 0) coordinate [circle,fill,inner sep=1pt] (pm1)
      (2, 0) coordinate [circle,fill,inner sep=1pt] (pp2)
      (-2, 0) coordinate [circle,fill,inner sep=1pt] (pm2)
      (3, 0) coordinate [circle,fill,inner sep=1pt] (pp3)
      (-3, 0) coordinate [circle,fill,inner sep=1pt] (pm3);

      \draw[ultra thin,gray] (hpoles) -- (p0) (hpoles) -- (pp1) (hpoles) --
(pm1) (hpoles) -- (pp2) (hpoles) -- (pm2) (hpoles) -- (pm3) (hpoles) -- (pp3);
    \end{tikzpicture}
  \end{center}
  \caption{Contour $\gamma$, enclosing the poles of $1/(1-\ee{-2\pi i t/h})$ on
    the real axis.}
  \label{fig:gamma}
\end{figure}

Consider the following contour integral over the rectangle $\gamma$ with corners
$-R-\ipih$, $R-\ipih$, $R+\ipih$ and $-R+\ipih$ (see \figref{gamma})
\begin{equation}
  \label{eq:C}
  C = \int_\gamma \frac{K(t;z)\,\ee{-t^2}\,\dd t}{1-\ee{-2\pi i t/h}}\,,
\end{equation}
where $R$ is real and positive. The function $1/(1-\ee{-2\pi i t/h})$ has simple
poles at all integer multiples of $h$ on the real axis, each with residue
$h/2\pi i$. Introducing it allows for the trapezoidal rule's formula to emerge
naturally from contour integration. Taking the limit $R\to+\infty$, the value of
$C$ can be computed by invoking the residue theorem. Let $z_k$, $k\in\mathbb{N}$
denote the poles of $K(t;z)$ lying inside the contour. Then the result can be
written as
\begin{equation}
  \label{eq:C_residues}
  C = h\sum_{n=-\infty}^{+\infty}K(nh;z)\,\ee{-n^2h^2} +
  2\pi i\sum_{k}\Res\left[\frac{K(t;z)\,\ee{-t^2}}
    {1-\ee{-2\pi i t/h}}\right]_{t=z_k}\,.
\end{equation}
The first term of \eqref{C_residues} takes into account the contribution from
the nodes of $1/(1-\ee{-2\pi i t/h})$. The second term corresponds to the
residues of the integrand at the poles $z_k$ of $K(t;z)$.

On the other hand, the integral can also be expressed as
\begin{equation}
  \label{eq:C_direct}
  C = \int_{-\infty-\ipih}^{+\infty-\ipih} \frac{K(t;z)\,\ee{-t^2}\,\dd t}{1-\ee{-2\pi i t/h}}
  + \int_{+\infty+\ipih}^{-\infty+\ipih} \frac{K(t;z)\,\ee{-t^2}\,\dd t}{1-\ee{-2\pi i t/h}}\,,
\end{equation}
where the contribution of the vertical segments located at $\real(t)=\pm R$
vanishes for $R\to+\infty$ due to the fast decay of the Gaussian term. The first
term on the right hand side can be rearranged to yield:
\begin{align}
  \label{eq:C_direct}
  C + \int_{-\infty+\ipih}^{+\infty+\ipih} \frac{K(t;z)\,\ee{-t^2}\,\dd t}{1-\ee{-2\pi i t/h}} =
  &\int_{-\infty-\ipih}^{+\infty-\ipih} K(t;z)\,\ee{-t^2}\,\dd t\\\nonumber
  +&\int_{-\infty-\ipih}^{+\infty-\ipih} \frac{K(t;z)\,\ee{-t^2-2\pi i
     t/h}\,\dd t}{1-\ee{-2\pi i t/h}}\,.\nonumber
\end{align}
Note that this manipulation is valid as long as the denominator does not vanish,
which is guaranteed over the line $\imag(t)=-\pi/h$ since the zeros of the
denominator are located on the real axis.

\begin{figure}[h]
  \begin{center}
    \begin{tikzpicture}[thick]
      \draw [->] (-4.5,0) -- (4.7,0) node [above left] {$\real(t)$};
      \draw [->] (0,-2) -- (0,0.7) node [left] {$\imag(t)$};

      \node (R) at (1.75, +0.4) [font=\scriptsize] {$R$};
      
      \draw [decorate, decoration={calligraphic brace}] (0.05,+0.1) -- (3.45,+0.1);

      \draw[green!60!black,decoration={markings,mark=between positions 0.1 and 1
step 0.25 with \arrow{>}},postaction={decorate}] (3.5,-1) -- (3.5,0) node [below
right] {$\gamma_-$} (3.5,0) -- (-3.5,0) (-3.5,0) -- (-3.5,-1) (-3.5,-1) --
(3.5,-1);

      \node (mih) at (-4.3, -0.5) [font=\scriptsize] {$\ipih$};
      \draw [decorate, decoration={calligraphic brace}] (-3.65,-0.9) -- (-3.65,-0.1);
    \end{tikzpicture}
  \end{center}
  \caption{Contour $\gamma_-$, overlapping with the real axis.}
  \label{fig:gammam}
\end{figure}
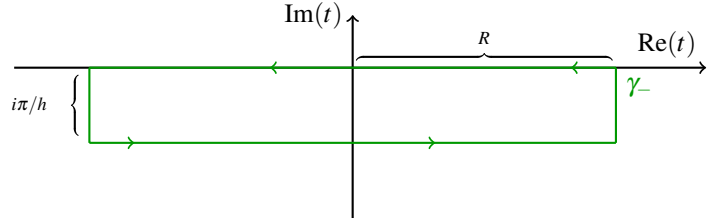

Now consider a second contour integral of the form
\begin{align}
  \label{eq:Cm_direct}
  C_- = \int_{\gamma_-} K(t;z)\,\ee{-t^2}\,\dd t =
  &\int_{-\infty-\ipih}^{+\infty-\ipih} K(t;z)\,\ee{-t^2}\,\dd t\\\nonumber
  +&\int_{+\infty}^{-\infty} K(t;z)\,\ee{-t^2}\,\dd t\,,\nonumber
\end{align}
where $\gamma_-$ has corners $-\infty-\ipih$, $+\infty-\ipih$, $+\infty+\ipih$
and $-\infty+\ipih$ (see \figref{gammam}, where again $R\ra+\infty$), and in
this case too the vertical segments do not contribute to the final value. Note
that the last term of this expression equals $-f(z)$ by definition. According to
the residue theorem, the value $C_-$ can also be computed as
\begin{equation}
  \label{eq:Cm_residues}
  C_- = 2\pi i\sum_{j}\Res\left[K(t;z))\,\ee{-t^2}\right]_{t=z_j}\,,
\end{equation}
where now $z_j$ are the poles of $K(t;z)$ lying in the region
$-\pi/h<\imag(t)<0$.

Combining \eqref{Cm_direct} and \eqref{Cm_residues}, one finds that
\begin{equation}
  \label{eq:Cm_final}
  f(z) = -2\pi i\sum_{z_j}\Res\left[K(t;z)\,\ee{-t^2}\right]_{t=z_j}
  + \int_{-\infty-\ipih}^{+\infty-\ipih} K(t;z)\,\ee{-t^2}\,\dd t\,.
\end{equation}
At this point, it can be noticed that the integral term in the right hand side
of \eqref{Cm_final} is the same as the first term in the right hand side of
\eqref{C_direct}, allowing $f(z)$ to be expressed as:
\begin{align}
  \label{eq:C_unevaluated}
  f(z) = &-2\pi i\sum_{z_j}\Res\left[K(t;z))\,\ee{-t^2}\right]_{t=z_j}+ C\\\nonumber
  &+\int_{-\infty+\ipih}^{+\infty+\ipih} \frac{K(t;z)\,\ee{-t^2}\,\dd t}{1-\ee{-2\pi i t/h}}
  -\int_{-\infty-\ipih}^{+\infty-\ipih} \frac{K(t;z)\,\ee{-t^2-2\pi i t/h}\,\dd t}{1-\ee{-2\pi i t/h}}\,.
\end{align}

Finally, we plug in the value of $C$ from \eqref{C_residues}. The expression to
evaluate $f(z)$ using the trapezoidal rule therefore reads
\begin{align}
  \label{eq:trapezoidal1}
  f(z) = &h\sum_{n=-\infty}^{+\infty}K(nh;z)\,\ee{-n^2t^2}
           +2\pi i\sum_{z_i}\Res\left[\frac{K(t;z)\,\ee{-t^2}}{1-\ee{-2\pi i t/h}}\right]_{t=z_i}\\\nonumber
  -&2\pi i\sum_{z_j}\Res\left[K(t;z)\,\ee{-t^2}\right]_{t=z_j} - E(h;z)\,,
\end{align}
where the error term $E(h;z)$ is given by the two last terms of
\eqref{C_unevaluated}. Exploiting the evenness of $K(t;z)$ and $\ee{-t^2}$ with
respect to $t$ this expression can be simplified. The final formula reads
\begin{align}
  \label{eq:trapezoidal}
  f(z) = & hK(0;z) + 2h\sum_{n=1}^{+\infty}K(nh;z)\,\ee{-n^2t^2}
           +2\pi i\sum_{z_i}\Res\left[\frac{K(t;z)\,\ee{-t^2}}{1-\ee{-2\pi i t/h}}\right]_{t=z_i}\\\nonumber
  -&2\pi i\sum_{z_j}\Res\left[K(t;z)\,\ee{-t^2}\right]_{t=z_j} - E(h;z)\,.
\end{align}

Note that in the derivation above I have assumed for simplicity of notation that
$K(t;z)$ has no poles falling on the lines $\imag(t)=\pm\ipih$ or $\imag(t)=0$.
Should this happen instead, and assuming these poles to be of order $1$ (which
is the case for the present work, and possibly the most relevant in general),
their contribution would be halved with respect to the formulas stated above.
Note that if such poles occur on the real axis, \eqref{f_integral} should be
interpreted as a Cauchy principal part.

The error term $E(h;z)$ of \eqref{trapezoidal} reads
\begin{equation}
  E(h;z) = +\int_{-\infty-\ipih}^{+\infty-\ipih} \frac{K(t;z)\,\ee{-t^2-2\pi i t/h}\,\dd t}{1-\ee{-2\pi i t/h}}
  -\int_{-\infty+\ipih}^{+\infty+\ipih} \frac{K(t;z)\,\ee{-t^2}\,\dd t}{1-\ee{-2\pi i t/h}}\,.
\end{equation}
Written in this way it is not especially illuminating or useful. However it can
be manipulated into a much clearer form. First, change the integration variable
as $t\to -t$ in the second term (which leaves $K$ and the Gaussian term
unaffected since they are even) to find
\begin{equation}
  E(h;z) = 2\int_{-\infty-\ipih}^{+\infty-\ipih}
  \frac{K(t;z)\,\ee{-t^2-2\pi i t/h}}{1-\ee{-2\pi i t/h}}\,\dd t\,.
\end{equation}
By the substitution $t=y-\ipih$ this is then written as
\begin{equation}
  E(h;z) = 2\int_{-\infty}^{+\infty}
  \frac{K(y-\ipih;z)\,\ee{-y^2}\,\ee{-\pi^2/h^2}}{1-\ee{-2\pi i
    y/h}\,\ee{-2\pi^2/h^2}}\,\dd y\,,
\end{equation}
where the integration is now on the real axis. The term $\ee{-2\pi^2/h^2}$ in
the denominator will be negligible with respect to 1 in all cases of practical
interest, so the expression simplifies to
\begin{equation}
  \label{eq:Ehz}
  E(h;z) = 2\ee{-\pi^2/h^2}\int_{-\infty}^{+\infty}
  K(y-\ipih;z)\,\ee{-y^2}\,\dd y\,.
\end{equation}
As noted by \cite{Goodwin1949}, as long as $K(t;z)$ does not grow too fast as
$|t|\ra+\infty$, the error can be approximated satisfactorily by setting
$K(t;z)\approx K(0;z)$ in \eqref{Ehz}, which results in the simple but useful
estimate:
\begin{equation}
  \label{eq:Ehz_approx}
  E(h;z) = 2\sqrt{\pi}\ee{-\pi^2/h^2}K(\ipih;z)\,.
\end{equation}

In many applications, such as the present work, it can be beneficial to have the
nodes of the trapezoidal rule on the real axis staggered with respect to 0, \ie
located at $h/2$, $3h/2$, $5h/2$, \dots instead of $0$, $h$, $2h$, \dots
\cite{Hunter1972}. This becomes necessary if some pole of $K(t;z)$ coincides
with one of the poles of $1/(1-\ee{-2\pi i t/h})$, since in this case expression
\eqref{trapezoidal} will become singular. This is easily achieved by considering
instead of \eqref{C} the integral
\begin{equation}
  \label{eq:Cstag}
  C_{\tn{staggered}} = \int_\gamma \frac{K(t;z)\,\ee{-t^2}\,\dd t}{1+\ee{-2\pi i t/h}}\,,
\end{equation}
where the contour $\gamma$ is the same as above, and the function
$1/(1+\ee{-2\pi i t/h})$ has simple poles at $t=(n-1/2)h$ for $n\in\mathbb{Z}$,
each with residue $h/2\pi i$. It is easy to see that the calculations above can
be carried on in the same fashion, resulting in the final formula
\begin{align}
  \label{eq:trapezoidals}
  f(z) = & 2h\sum_{n=1}^{+\infty}K\left[(n-1/2)\,h;z\right]\,\ee{-(n-1/2)^2t^2}
           +2\pi i\sum_{z_i}\Res\left[\frac{K(t;z)\,\ee{-t^2}}
           {1+\ee{-2\pi i t/h}}\right]_{t=z_i}\\\nonumber
  -&2\pi i\sum_{z_j}\Res\left[K(t;z)\,\ee{-t^2}\right]_{t=z_j} + E(h;z)\,,
\end{align}
where the error term is still given by \eqref{Ehz}.

\bibliographystyle{plain}
\bibliography{refs}

\end{document}